\documentclass[showpacs,aps,
twocolumn,showpacs,preprintnumbers,superscriptaddress]{revtex4-1}
\usepackage{amsmath}
\usepackage{graphicx}
\usepackage{dcolumn}
\usepackage{bm}
\usepackage{color}
\usepackage{epstopdf}

\graphicspath{{figs/}}

\begin{document}


\title{Force-induced transparency and conversion between slow and fast lights
in optomechanics}
\author{Zhen Wu}
\affiliation{City College, Wuhan University of Science and Technology, Wuhan 430083, China}
\author{ Ren-Hua Luo}
\affiliation{ Hubei Electric Engineering Corporation (POWERCHINA HEEC), Wuhan 430040, China}
\author{Jian-Qi Zhang}
\email{changjianqi@gmail.com}
\affiliation{State Key Laboratory of Magnetic Resonance and Atomic and Molecular Physics, Wuhan Institute of Physics and Mathematics, Chinese Academy of Sciences, Wuhan 430071, China}
\author{Yu-Hua Wang}
\affiliation{Hubei province Key Laboratory of Science in Metallurgical Process, Wuhan University of Science and Technology, Wuhan 430081, China}
\author{Wen Yang}
\affiliation{Beijing Computational Science Research Center, Beijing 100084, China }
\author{Mang Feng}
\email{mangfeng@wipm.ac.cn}
\affiliation{State Key Laboratory of Magnetic Resonance and Atomic and Molecular Physics, Wuhan Institute of Physics and Mathematics, Chinese Academy of Sciences, Wuhan 430071, China}
\affiliation{Department of Physics, Zhejiang Normal University, Jinhua 321004, China }
\date{\today }

\begin{abstract}
The optomechanics can generate fantastic effects of optics due to
appropriate mechanical control. Here we theoretically study effects of slow
and fast lights in a single-sided optomechanical cavity with an external
force. The force-induced transparency of slow/fast light and the
force-dependent conversion between the slow and fast lights are resulted
from effects of the rotating-wave approximation (RWA) and the anti-RWA, which can be controlled
by properly modifying the effective cavity frequency due to the
external force. These force-induced phenomena can be applied to control of
the light group velocity and detection of the force variation, which are
feasible using current laboratory techniques.
\end{abstract}
\maketitle

\section{Introduction}
The cavity optomechanics (COM), combining mechanical modes with optical
modes via radiation pressure, has attracted considerable attention recently.
Extensive research efforts have presented interesting quantum properties and
nonlinear effects by optomechanics, such as entanglement \cite%
{Bose1997Preparation,Vitali2007Optomechanical,Sun2012First}, squeezing \cite%
{Liao2011Parametric,Han2013The}, normal mode splitting \cite{Huang2009Normal}%
, Kerr effect \cite{Xin2013Quantum}, optomechanically induced transparency
(OMIT) \cite{agarwal2010electromagnetically,Huang2011Electromagnetically,
Safavinaeini2011Electromagnetically,weis2010optomechanically,Bai2016Tunable},
optical solitons\cite{Gan2016Solitons}, and chaos\cite%
{L2015PT,Bakemeier2014Route}, which are associated with potential
applications in quantum information processing \cite%
{Agarwal2012Optomechanical,Xiong2015Asymmetric} and precision measurements
\cite{zhang2012precision,Aspelmeyer2014Cavity,He2015Sensitivity}.

Among the above mentioned items, the OMIT, a kind of induced transparency
arisen from the interference of excitation pathways in optomechanical
systems is the research focus of the present paper. We have noticed that
the electromagnetically induced transparency (EIT) in atoms can
produce slow and fast lights \cite{Hau1999Light}, which is the
technique with appealing applications in optical storage \cite%
{Tucker2006Slow,Xing2007Storage}, optical telecommunication \cite%
{Boyd2009Controlling}, signal processing \cite%
{Willner2008Optical} and interferometry \cite{Shi2007Enhancing}. So we
wonder if the OMIT, with analogy to the EIT, could also work for producing
slow/fast light and even beyond. In fact, there have been publications
\cite{Chen2011Slow,zhan2013tunable,tarhan2013superluminal,jiang2013electromagnetically,
tarhan2013superluminal1,akram2015tunable,akram2015Efficient}
for slow and fast light effects associated with the COM using similar behavior
to those with multi-level atoms. As shown below, however, we will go for a
further step with the COM by presenting an experimentally feasible proposal
for a force-induced transparency with slow/fast light and a conversion
between the slow and fast lights.

Specifically, different from the traditional transparency proposals \cite%
{Imamo1991Observation,Fleischhauer2005Electromagnetically, apl-109-261106,Nature-465-755},
where the slow/fast light can only be adjusted with an external optical field, e.g.
the power and frequency of the pump field, our study shows
that we can control exactly by an external force the group velocity of lights with a fixed
pump field. This external force employed for the control could be
Coulomb-relevant \cite{zhang2012precision} or magnetic effects \cite{prl-108-120801}.

Two kinds of external forces are usually employed in the optomechanics. One is the constant force \cite{nanotech-7-509,prl-114-093602} including
electric field force \cite{zhang2012precision}, magnetic field force \cite{prl-108-120801}, elastic force \cite{apl-105-014108,pra-94-043855} and optical gradient force \cite{nphoton-4-211};
The other is the time-harmonic-driving force \cite{pra-91-043843,srep-5-11278}, which could be
achieved with piezoelectric coupling \cite{nphys-9-712} and Lorentz force \cite{njp-9-35}.
In our scheme, we choose a constant force as the external force, which can modify the eigen-frequencies of the cavity by adjusting the cavity length.
Such an external force is similar to the one in Ref.\cite{prl-114-093602}, where the force
is applied to balance the effective force from the nonlinear optical effect.

Compared with the optical manipulations on the group velocity of lights \cite%
{Imamo1991Observation,Fleischhauer2005Electromagnetically, apl-109-261106,Nature-465-755},
which are limited by the power and frequency ranges of the laser field, the external force in our scheme works
in a larger regime. Due to the external force, the effective eigen-frequency of the optomechanical
cavity is modified, and thus we may achieve the conversion between slow and fast
lights in this way. This conversion is physically governed by the conditions for the
rotating-wave approximation (RWA) and the anti-RWA
which are present as anti-Stokes and Stokes processes in the optomechanics, respectively.

As pointed out below, the slow/fast light is originally from
the effect of RWA/anti-RWA of the parameters in the system, which is
a more fundamental factor than the anti-Stokes/Stokes process as mentioned in Refs. \cite{akram2015tunable,akram2015Efficient}.
The latter is valid only for the situation of the third-order nonlinear coupling, but the former can explain the fast/slow light effects in various
physical systems including both the linear \cite{Imamo1991Observation,Fleischhauer2005Electromagnetically,apl-109-261106,Nature-465-755}
and the nonlinear coupling systems \cite{Chen2011Slow,zhan2013tunable,tarhan2013superluminal,jiang2013electromagnetically,
tarhan2013superluminal1,akram2015tunable,akram2015Efficient}.
Moreover, our proposal is more simplified and
effective, and experimentally feasible using current techniques \cite{nanotech-7-509}
since it is only required to apply an external force on the optomechanical
resonator, much more easily adjusted than the idea with an additional atom
\cite{akram2015tunable} or nano-resonator \cite{akram2015Efficient}.
In Ref.\cite{akram2015tunable}, the slow/fast light is adjusted by the detuning
between the optomechanical cavity and a cavity-confined atom, which is hard to manipulate
experimentally. The control of the slow/fast light in \cite{akram2015Efficient} is made
by Coulomb coupling between two nanomechanical resonators (NRs), which is also experimentally challenging.

Furthermore, the effects in our work can be observed even
in room temperature since the noise is much less than the mean value of the output field \cite{pra-88-013804}.
In particular, due to one-to-one correspondence between the external force
and the group velocity of the light in some special regimes, our proposal
could be used to achieve precision measurements and operations. As such, our
work provides a new and effective way to control the group velocity of the
light in optomechanical systems with an external force.

The rest of the paper is structured as follows. In Sec. II, we present
the Hamiltonian and the steady state of the optomechanics. In Sec.
III, we deduce the output of the probe field and its time delay. In Sec. IV, we give
some simulations and discussions for the force-induced light
transparency and the force-dependent conversion between the slow and
fast lights under some experimentally available conditions. The conclusion is given in the last section.

\section{Hamiltonian and Steady states}

\begin{figure}[tbp]
\includegraphics[width=6cm]{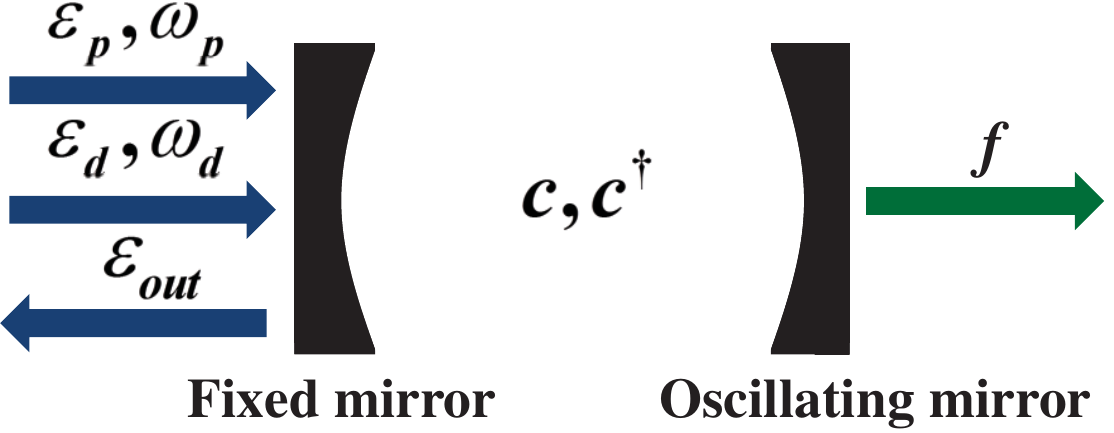}\newline
\caption{Schematic diagram of the system. The COM consists of a fixed mirror
and an oscillating mirror. The arrow with labels $\protect\varepsilon_{p}$%
, $\protect\omega_{p}$ is for a weak probe field and the arrow with labels
$\protect\varepsilon_{d}$, $\protect\omega_{d}$ for a strong pump field.
The output field is with the field amplitude $\protect\varepsilon_{out}$. An
external force is applied on the oscillating mirror for producing the
slow/fast light effects as discussed in the text.}
\label{model}
\end{figure}

As sketched in Fig. \ref{model}, an optomechanical cavity is driven by a
strong pump field with frequency $\omega _{d}$ and power $P_{d}$, and by a weak
probe field with frequency $\omega _{p}$ and power $P_{p}$. The oscillating
mirror with mass $m$ and frequency $\omega _{m}$ couples to the Fabry-Perot
cavity with frequency $\omega _{c}$ via radiation pressure force, and
experiences an external force $f$. As mentioned above, this force must be a constant force, such as
a Coulomb force \cite{zhang2012precision} or a magnetic force with a steady electric current \cite{Ma2015Optomechanically}.

In the rotating frame at frequency $\omega _{d}$ of the pump field, the
Hamiltonian of the system is given by \cite%
{zhang2012precision,Ma2015Optomechanically}
\begin{equation}
\begin{split}
H=& \hbar \Delta _{c}c^{\dagger }c+(\frac{p^{2}}{2m}+\frac{1}{2}m\omega
_{m}^{2}q^{2})-\chi qc^{\dagger }c-fq \\
& +i\hbar \lbrack (\varepsilon _{d}+\varepsilon _{p}e^{-i\delta
t})c^{\dagger }-H.c.],
\end{split}
\label{Ham}
\end{equation}
with detunings $\Delta _{c}=\omega _{c}-\omega _{d}$ and $\delta =\omega
_{p}-\omega _{d}$. The first term is the free Hamiltonian for the cavity
with the annihilation (creation) operator $c$ $(c^{\dagger})$. The second term
describes the energy for the oscillating mirror with $q$ ($p$) being the
position (momentum) operator. The third term represents the radiation
pressure effect between the cavity and the oscillating mirror with a
coupling strength $\chi =\frac{\hbar \omega _{c}}{L}$, where $L$ is the
cavity length. The forth term is associated with the external force on the
oscillating mirror. The last two terms are the interactions between the
cavity and two input fields with strengths $\varepsilon _{d}=\sqrt{\frac{%
2\kappa P_{d}}{\hbar \omega _{d}}}$ and $\varepsilon _{p}=\sqrt{\frac{%
2\kappa P_{p}}{\hbar \omega _{p}}}$, respectively, where $\kappa$ is cavity
decay rate.

To get the mean response of the system, we employ the Heisenberg-Langevin
equations and the mean-field approximation \cite%
{agarwal2010electromagnetically}. Then the mean-value equations of our model
can be written as
\begin{equation}  \label{MV-eq}
\begin{aligned} \langle\dot{q}\rangle &= \frac{\langle p \rangle}{m}, \\
\langle\dot{p}\rangle &= -m\omega_m^2\langle q\rangle+\chi\langle c^\dagger
c\rangle+f-\gamma_m\langle p\rangle, \\ \langle\dot{c}\rangle
&=-[\kappa+i(\Delta_c-\frac{\chi}{\hbar}\langle q\rangle)]\langle
c\rangle+\varepsilon_d+\varepsilon_pe^{-i\delta t}, \end{aligned}
\end{equation}
where $\gamma_{m}$ is the decay rate of the movable mirror. The steady-state response of Eq. (\ref{MV-eq})
contains many Fourier components, where we are only interested in the linear response of
the system for the probe field.

\begin{figure}[tbp]
\centering
\includegraphics[width=8.5cm]{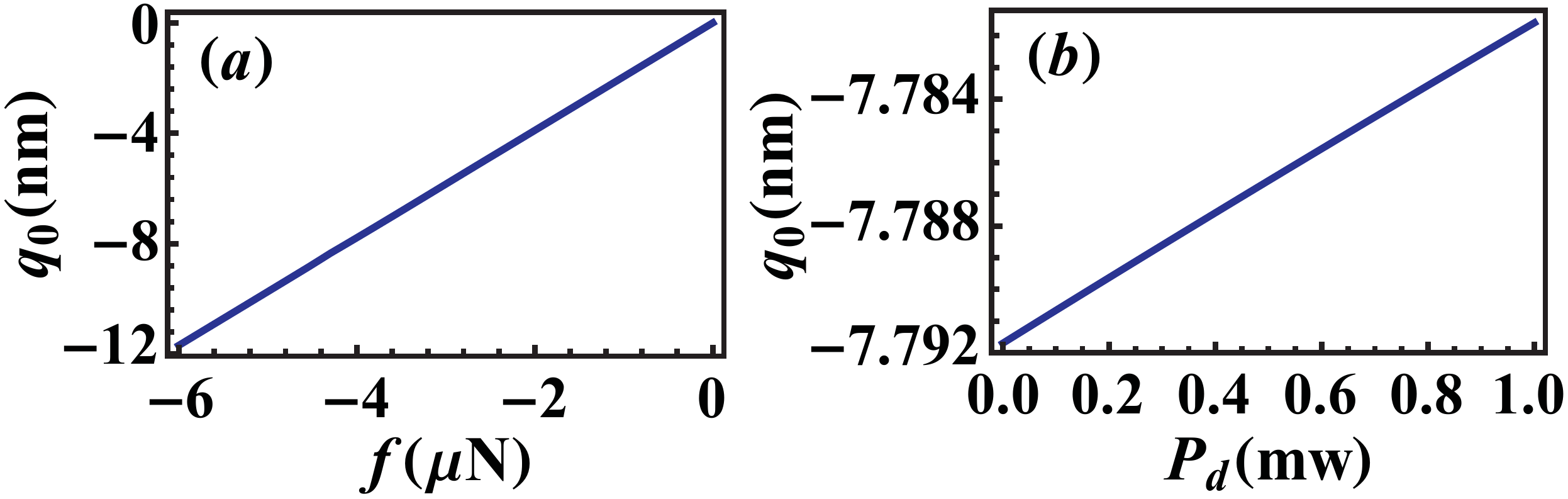}
\caption{(a) The steady state for the position $q_{0}$ versus the external force
$f$ in the case of the pump field power $P_{d}=0.2$ mW. (b) The steady state for the position
$q_{0}$ as a function of the pump field power in the case of the external force $f=-4\times10^{-6}$ N.
Other parameters used are $\protect\omega_{d}=2\protect\pi c/\protect\lambda$ with
$\protect\lambda =1064$ nm, $\Delta_c=-10 \protect\omega_m$, $\protect\kappa =2\protect\pi\times
215$ kHz, $m=145$ ng, $\protect\omega_{m}=2\protect\pi\times 947$ kHz, $%
\protect\gamma_{m}=2\protect\pi\times 141$ Hz.}
\label{compared}
\end{figure}

To obtain steady-state solutions to Eq. (\ref{MV-eq}), we assume the
equation \cite{Boyd2008Quantum} $\langle s\rangle =s_{0}+s_{+}\varepsilon
_{p}e^{-i\delta t}+s_{-}\varepsilon _{p}^{\ast }e^{i\delta t}$ with $s=q,p,c$%
, and these three terms $s_{0,\pm}$ are associated
with the frequencies $\omega _{d}$, $\omega _{p}$, $2\omega _{d}-\omega _{p}$%
, respectively. Inserting the three equations into Eq. (\ref{MV-eq}), we
obtain
\begin{equation}
\begin{aligned} q_0 &= \frac{\chi |c_0|^2+f}{m\omega_m^2}, \quad c_0
=\frac{\varepsilon_d}{\kappa+i\Delta},\\
c_{+}&=\frac{(\delta^2-\omega_m^2+i\gamma_m\delta)[\kappa-i(\Delta+\delta)]-2i%
\omega_m\beta}
{(\delta^2-\omega_m^2+i\gamma_m\delta)[\Delta^2+(\kappa-i\delta)^2]+4\Delta%
\omega_m\beta}, \end{aligned}  \label{ss-eq}
\end{equation}%
where $\beta =\chi ^{2}|c_{0}|^{2}/(2m\hbar \omega _{m})$, and $\Delta
=\Delta _{c}-\chi q_{0}/\hbar $ is the effective cavity-pump detuning,
depending on the steady-state position $q_{0}$ of the mirror. Assuming that the above solutions are
based on the mean value much larger than the noise, we consider that the effects resulted from those
solutions could be observed at room temperature, similar to the one in Ref. \cite{pra-88-013804}.

With the steady-state solution in Eq. (\ref{ss-eq}), the steady-state
equation for the position $q_{0}$ can be rewritten as
\begin{equation}
\begin{array}{cc}
m\omega _{m}^{2}\frac{\chi ^{2}}{\hbar ^{2}}q_{0}^{3}-(f\frac{\chi ^{2}}{%
\hbar ^{2}}+2m\omega _{m}^{2}\frac{\chi }{\hbar }\Delta _{c})q_{0}^{2} &  \\
+[m\omega _{m}^{2}(\kappa ^{2}+\Delta _{c}^{2})+2f\frac{\chi }{\hbar }\Delta
_{c}]q_{0}-[f(\kappa ^{2}+\Delta _{c}^{2})+\chi \varepsilon _{d}^{2}] & =0,%
\end{array}
\label{ss-q0}
\end{equation}%
which means that the steady state for the position $q_{0}$ depends on two
tunable parameters: the pump power $\varepsilon _{d}$ and the external force
$f$ (see Fig. \ref{compared}). In other words, the effective cavity frequency (cavity-pump detuning) $%
\omega _{c}^{\prime }=\omega _{c}-\chi q_{0}/\hbar $ ($\Delta =\Delta
_{c}-\chi q_{0}/\hbar $) can be adjusted by controlling the pump power and
the external force. It reminds us the possibility to realize some
force-induced/dependent physics.

\section{Output light and time delay}

With the application of the input-output relation \cite{Walls1995Quantum} $%
\varepsilon _{out}=\varepsilon _{in}-2\kappa \langle c\rangle ,$ we have the
output field
\begin{equation}
\varepsilon _{out}=(\varepsilon _{d}-2\kappa c_{0})+(1-2\kappa
c_{+})\varepsilon _{p}e^{-i\delta t}-2\kappa c_{-}\varepsilon _{p}^{\ast
}e^{i\delta t}.  \label{output}
\end{equation}
For simplicity, we assume the quadrature of the output field as
\begin{equation}
\begin{array}{lll}
\varepsilon _{T} & = & 2\kappa c_{+} \\
& = & \dfrac{2\kappa }{[\kappa -i(\delta -\Delta )]+\frac{2i\omega _{m}\beta
}{(\delta ^{2}-\omega _{m}^{2}+i\gamma _{m}\delta )-\frac{2i\omega _{m}\beta
}{\kappa -i(\delta +\Delta )}}},
\end{array}
\label{quadrature}
\end{equation}
whose real and imaginary parts are associated with the absorption and dispersion,
respectively \cite{agarwal2010electromagnetically}. Moreover, the output
field varies with both the pump strength $\varepsilon_{d}$ and the external
force $f$, implying that the external force, in addition to the pump field, can construct the
light transparency (see Eq. (\ref{ss-q0}) and Fig. \ref{compared}).

To follow the force-induced transparency of the probe light, we suppose the
system working in the resolved-sideband regime due to $\omega_{m}\gg\kappa$. This
is the condition for the normal mode splitting in optomechanics, and the
strongest radiation coupling can be achieved when the system reaches the first-order red/blue sideband with $\delta =\pm \omega_{m}$ or $\delta =\pm\Delta$.

\begin{figure}[tbp]
\centering
\includegraphics[width=8.5cm]{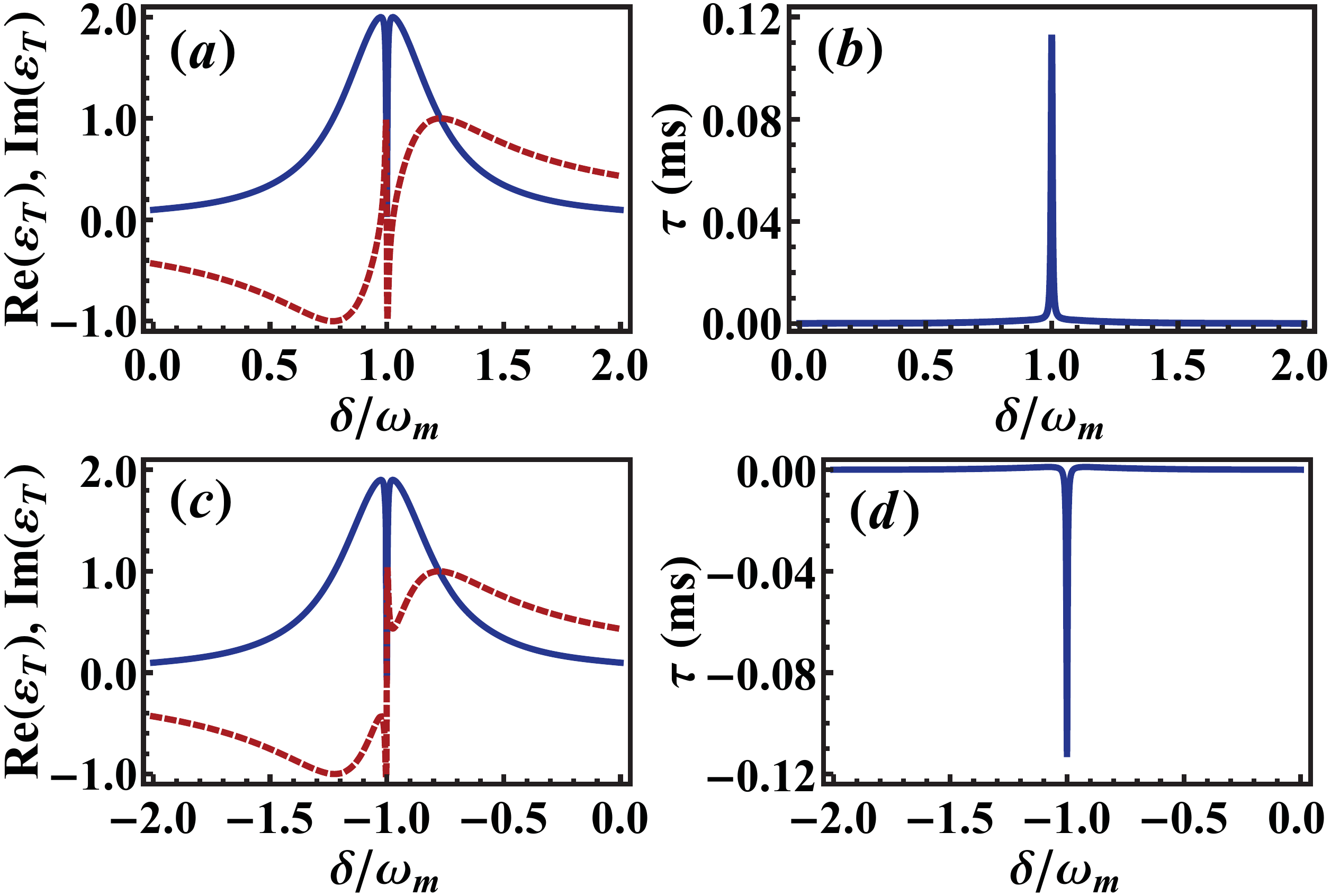}
\caption{(a) The real (solid line) and imaginary (dashed line) parts of $\protect\varepsilon_{T}$ as functions of $\delta/\omega_{m}$ with
$f_1=-4.74\times 10^{-6}$ N. (b) The corresponding group velocity delay $\protect\tau$
versus the normalized frequency $\protect\delta/\protect\omega_{m}$ for $f_1=-4.74\times 10^{-6}$ N.
(c) The real (solid line) and imaginary (dashed line) parts of $\protect\varepsilon_{T}$ as functions of $\delta/\omega_{m}$ for
$f_2=-3.88\times 10^{-6}$ N. (d) The corresponding group delay $\protect\tau$ versus the
normalized frequency $\protect\delta /\protect\omega_{m}$ for $f_2=-3.88\times 10^{-6}$ N.
Other parameters are the same as in Fig. \protect\ref{compared}.}
\label{Fig2_slow}
\end{figure}

In the case of $\Delta\simeq\omega_{m}$ (i.e., the RWA case), the optomechanics works in the first-order red sideband. With
the application of $\delta^{2}-\omega_{m}^{2}\simeq
2\omega_{m}(\delta -\omega_{m})$, we neglect the small term $2i\omega_{m}\beta
/[\kappa -i(\delta +\Delta)]$. Thus Eq. (\ref{quadrature}) is rewritten as
\begin{equation}
\begin{array}{lll}
\varepsilon _{T} & \simeq  & \dfrac{2\kappa }{[\kappa -i(\delta -\Delta )]+%
\dfrac{2i\omega _{m}\beta }{\delta ^{2}-\omega _{m}^{2}+i\gamma _{m}\delta }}
\\
& \simeq  & \dfrac{2\kappa }{[\kappa -i(\delta -\Delta )]+\dfrac{\beta }{\frac{%
\gamma _{m}}{2}-i(\delta -\omega _{m})}},%
\end{array}
\label{slowlight}
\end{equation}
which is the expression for the slow light [see Figs. \ref{Fig2_slow}(a) and (b)].

Similarly, in the case of $\Delta\simeq -\omega_{m}$ (i.e., the case of anti-RWA), the system is governed by the first-order blue sideband, and
we have $\delta^{2}-\omega_{m}^{2}\simeq -2\omega_{m}(\delta +\omega_{m})$,
\begin{equation}
\begin{array}{lll}
\varepsilon _{T} & \simeq  & \dfrac{2\kappa }{[\kappa -i(\delta -\Delta )]-%
\dfrac{\beta }{\frac{\gamma _{m}}{2}-i(\delta +\omega _{m})}},%
\end{array}
\label{fastlight}
\end{equation}%
which is the solution for the fast light [see Figs. \ref{Fig2_slow}(c) and (d)]. In this situation,
a very small gain can be achieved in the absorption of the output field ($Re[\varepsilon_{T}]$).
This gain of the probe light originates from the anti-RWA process with both a photon and a phonon simultaneously created or annihilated.
Due to the large cavity decay, however, the photon-phonon creation dominates the system evolution, which is supported by the external field.
As such, we have the gain in the absorption of the output field.

The slow/fast light conversion in our scheme is very different from the previously proposed transparencies using atoms \cite%
{Imamo1991Observation,Fleischhauer2005Electromagnetically}, coupled cavities
\cite{apl-109-261106} and atom-cavity hybrids \cite{Nature-465-755}.
Compared with Eq. (\ref{slowlight}), Eq. (\ref{fastlight}) has a sign difference in the denominator, which can be understood as
a switch between the effects of RWA and anti-RWA. Since the two solutions of $\varepsilon_{T}$ under
RWA and anti-RWA take two fast changes in absorption/dispersion ($Re[\varepsilon_{T}]$/$Im[\varepsilon_{T}]$)
with the slopes in different signs, the conversion between the slow and
fast lights can be achieved by controlling the parameters to reach the RWA and anti-RWA regimes.
This viewpoint is different from in Refs. \cite{akram2015tunable,akram2015Efficient},
where the fast/slow light is explained as a characteristic in the anti-Stokes/Stokes process.
Actually, the anti-Stokes/Stokes process \cite{akram2015tunable,akram2015Efficient} owns the fast/slow light due to the third-order nonlinear
coupling as the radiation coupling. In contrast, by the fact that Eq. (\ref{quadrature}) is reduced
to Eq. (\ref{slowlight}) under the RWA and to Eq. (\ref{fastlight}) under the anti-RWA,
we consider that the slow and fast effects originate fundamentally from the RWA and the anti-RWA employed
for the parameters. This is a more fundamental reason than the anti-Stokes/Stokes effects for the slow/fast light in general systems.
For example, the slow light effect observed in the previous publications
\cite{Imamo1991Observation,Fleischhauer2005Electromagnetically,apl-109-261106,Nature-465-755} is due to the involvement of only the RWA.


\begin{figure}[tbp]
\centering
\includegraphics[width=8.0cm]{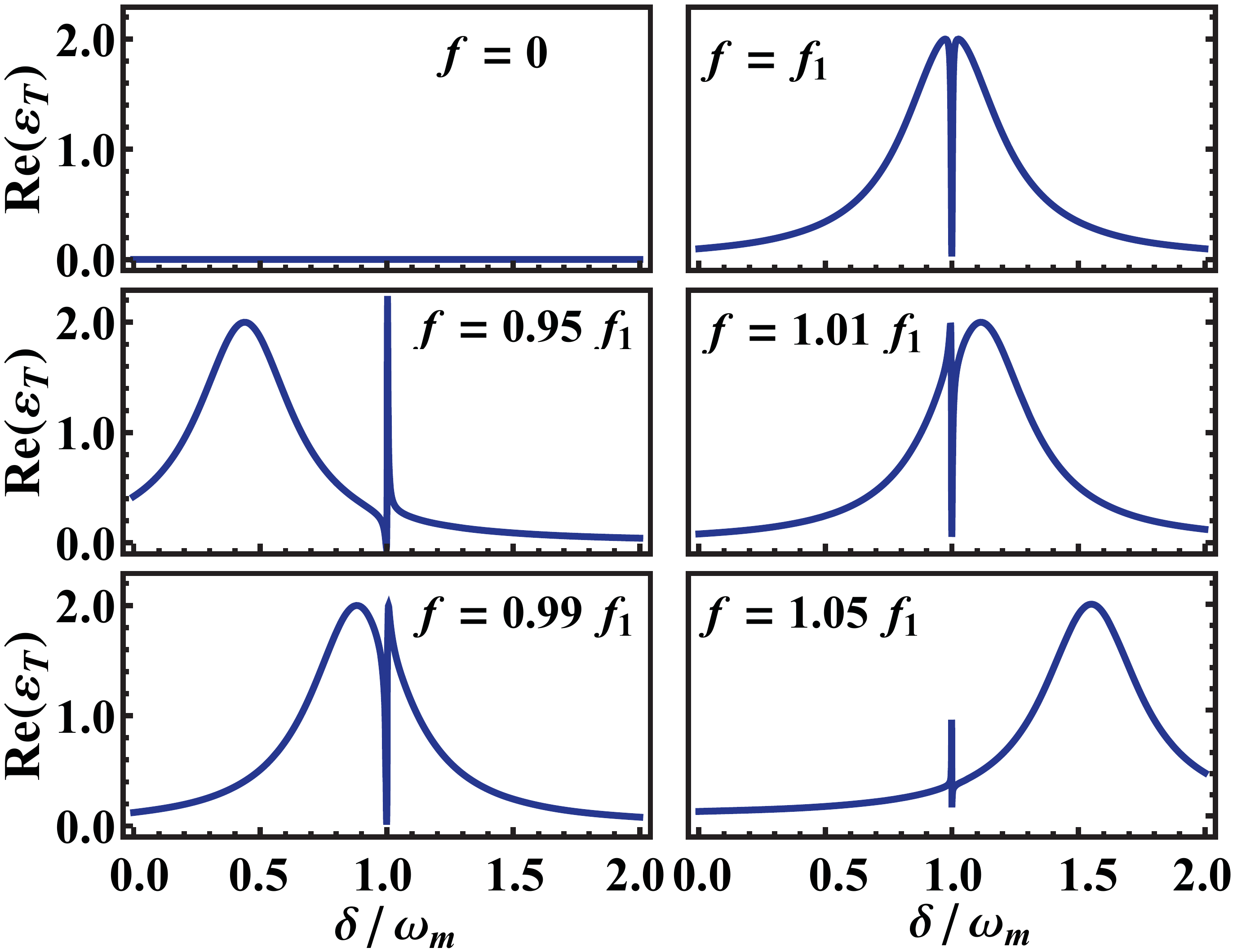}
\caption{The real parts of $\protect\varepsilon _{T}$ as functions of the
force $f$ under the condition of RWA, where we consider six typical values of $f/f_{1}$ from zero to
a value larger than 1.0. Other parameters are the same as in Fig. \protect\ref%
{Fig2_slow}.}
\label{Fig3_force-OMIT}
\end{figure}

The transmission of the probe field, defined by the ratio of the output and
input field amplitudes at the frequency of the probe light \cite%
{weis2010optomechanically,zhan2013tunable}, is given by
\begin{equation}
\varepsilon =1-2\kappa c_{+}.  \label{trans}
\end{equation}%
In the regime of the narrow transparency window, there is a rapid variation
of the probe phase $\Phi (\omega _{p})=\mathrm{arg}[\varepsilon ]=\frac{1}{2i%
}\mathrm{ln}(\frac{\varepsilon }{\varepsilon ^{\ast }})$. This variation is
associated with the group velocity delay as \cite{tarhan2013superluminal1,gu2015tunable}
\begin{equation}
\tau =\frac{\partial\Phi}{\partial\omega_{p}}|_{\bar{\omega}}=\mathrm{Im}[%
\frac{1}{\varepsilon}\frac{\partial\varepsilon}{\partial\omega_{p}}]|_{\bar{%
\omega}}=\mathrm{Im}[\frac{1}{\varepsilon}\frac{\partial\varepsilon }{%
\partial\delta}]|_{\delta=\pm\omega_{m}},  \label{groupdelay}
\end{equation}%
where $\bar{\omega}=\omega_{d}\pm\omega_{m}$, and $\delta
=\pm\omega_{m}$ is the condition for the two-photon resonance. $\tau>0$ and $%
\tau<0$ correspond to the slow and fast light propagation, respectively.

\section{Simulations and discussion}

To demonstrate the force-induced light transparency and the force-dependent
conversion between the slow and fast lights, we have made some simulations
using following experimental parameters \cite{groblacher2009observation}:
$\lambda =1064$ nm, $P_{d}=0.2$ mW,
$L=25$ mm, $\kappa /2\pi =215$ kHz, $m=145$ ng, $\omega _{m}/2\pi =947$ kHz,
$\gamma _{m}/2\pi =141$ Hz. In what follows, to justify the feasibility of our scheme,
we assume a fixed pump light which is far detuned from the cavity as $\Delta
_{c}=-10\omega_{m}$.

\subsection{Force-induced light transparency}

In the absence of the external force, the effective detuning between the optomechanical cavity and the
pump field is $\Delta\approx -10\omega_{m}$, which is far detuned from the
resonator frequency $\omega _{m}$. In this case, even if the pump power is set to some feasible values, no
OMIT can be observed due to the large detuning of the pump light from the
cavity field and the limitation of the work regime for the optical manipulation.
However, if an external force $f=f_{1}$ is applied on the optomechanics, as shown in Fig. \ref{Fig2_slow},
the OMIT appears since the condition of $\Delta=\omega_{m}$ can be satisfied. This is
due to the fact that the external force pushes the system into the
red-sideband regime under the RWA by increasing the effective cavity frequency.
As shown in Fig. \ref{Fig3_force-OMIT}, with the increase of the force, the central frequency for
transparency remains unchanged, whereas the main peak of the output
field moves from the low frequency to the higher due to the increase of the
effective cavity frequency.

In contrast, with an alternative external force $f=f_{2}$ applied, the system moves
into the blue-sideband regime under the anti-RWA.
The fast light due to force-induced transparency is thus produced.

\begin{figure}[tbp]
\centering
\includegraphics[width=6.5cm]{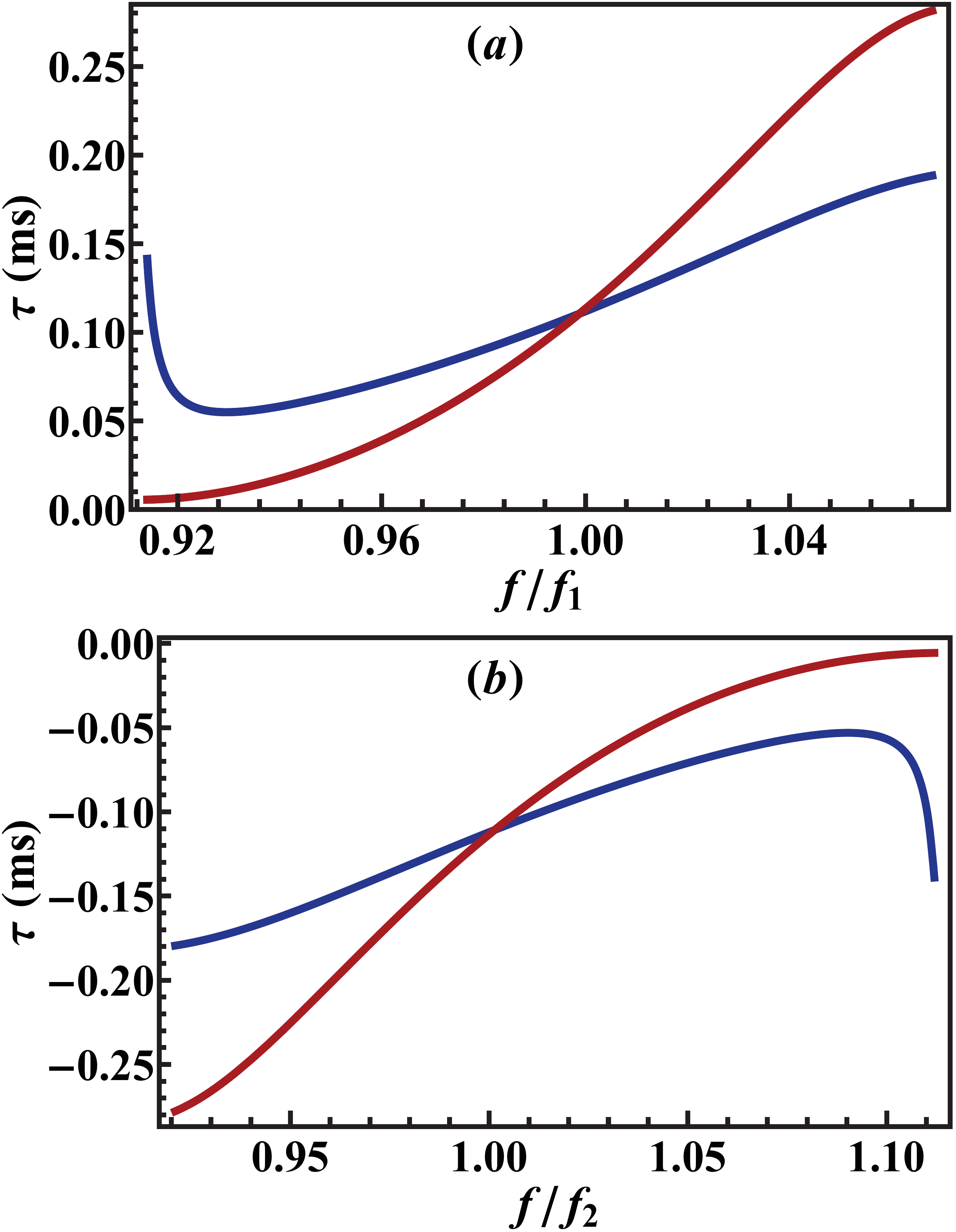}
\caption{(a) The group velocity delay as a function of $f/f_{1}$ with $f_{1}=-4.74\times 10^{-6}$ N and $\protect\delta =%
\protect\omega _{m}$. The blue (red) line is for an analytical (approximate) result by  Eq. (\ref{quadrature}) (Eq. (\ref{slowlight})). (b) The group velocity delay as a function of $f/f_{2}$ with  $f_{2}=-3.88\times 10^{-6}$ N and $\protect\delta =-%
\protect\omega_{m}$. The blue (red) line is for an analytical (approximate) result by Eq. (\ref{quadrature}) (Eq. (\ref{fastlight})). Other parameters are the same as in Fig. \protect\ref{Fig2_slow}.}
\label{Fig5_force-delay}
\end{figure}

\subsection{Force-dependent slow/fast light conversion}

As discussed above, when the external force $f=f_{1}$ is applied, there is a force-induced
transparency for the slow light [see Figs. \ref{Fig2_slow}(b)]. In contrast,
when the external force is $f=f_{2}$, the system works in the blue-sideband regime
with an effective detuning $\Delta =-\omega_{m}$, and the fast light effect
is available in this situation [see Figs. \ref{Fig2_slow}(d)]. In this context, it is natural to
ask if there is a possibility to have a conversion between the slow light and
the fast one.

The answer to this possibility is positive, as shown below. The physics for the control of the slow and fast lights can be understood
from Eq. (\ref{quadrature}). There are fast changes in absorption/dispersion in very
narrow spectral ranges [see Fig. \ref{Fig2_slow}(a) and (c)],
which are followed by large changes in the refractive index due to Eq. (\ref%
{quadrature}) satisfying the Kramers-Kronig relations \cite{OE-20-14009}. As
a result, if $f=f_{1}$, the system works in the red-sideband regime with
a positive change in the refractive index, yielding a low
group velocity [see Fig. \ref{Fig2_slow}(b)]. In contrast, the system turns to
be in the blue-sideband regime once $f=f_{2}$ is applied,
which creates a high group velocity [see Fig. \ref{Fig2_slow}(d)] due to a
negative change in the refractive index.

Since the probe transmission $\varepsilon$ depends on both the pump power
and the external force in our approach, the group velocity delay $\tau$ can be tuned by both
the power of the pump light and the external force, which are different from
previous ideas with $\tau$ modified only by the pump power \cite%
{tarhan2013superluminal,jiang2013electromagnetically,tarhan2013superluminal1,akram2015tunable}.
To show this, we plot $\tau$ as a function of the force around
the detuning of $\delta =\pm\omega_{m}$ in Fig. \ref{Fig5_force-delay}. It
implies that the external force $f$ can be used to control the group
velocity of the probe light even with a fixed pump field, and also means the
possibility to measure the external force using this property. In
particular, the delay $\tau$ approximately linearly varies with $f$ at
the point near $f=f_{1}$ ($f=f_{2}$) for which the slope is $d\tau
/df\approx$ 244(242) s/N at $\Delta\approx\omega_{m}$ ($\Delta
\approx -\omega_{m}$). Within the regime with one-to-one correspondence between the
group velocity and the external force, we may perform precision control or measurement for
the group velocity using a certain external force.

Moreover, in Fig. \ref{Fig5_force-delay}, with the increase of the external force $f$ for the slow (fast) light,
the approximate and analytic results intersect at the point of $f=f_{1}$ ($f=f_{2}$) where the system meets
exactly the condition for the red(blue)-sideband regime. In contrast to the previous works \cite{Chen2011Slow,tarhan2013superluminal,zhan2013tunable,jiang2013electromagnetically,
tarhan2013superluminal1,akram2015tunable,akram2015Efficient} with the time delay expressed by reduced analytic solutions,
we fully consider the contribution from the effects of both RWA and anti-RWA
in the measurement of the group velocity using a certain external force. In this context,
our work provides a further understanding of the slow and fast lights in comparison with
the previous treatments \cite{Chen2011Slow,tarhan2013superluminal,zhan2013tunable,jiang2013electromagnetically,
tarhan2013superluminal1,akram2015tunable,akram2015Efficient}.

\section{Conclusion}

In summary, we have studied and explained the slow and fast light effects in a
single-sided optomechanical cavity under an external force. The two special
characters of the optomechanical cavity, i.e., the force-induced light
transparency and conversion related to the slow/fast light, can be fully
controlled by the effective cavity frequency modified by the external force.
In particular, we pointed out that the effect of RWA/anti-RWA of the parameters
is the fundamental reason to generate the slow/fast lights. Since our proposal
 is feasible using current laboratory techniques,
we believe that our scheme provides a new way to producing
tunable fast and slow lights, which helps inspiring more potential
applications for optomechanics.

\begin{acknowledgments}
JQZ thanks Yong Li for helpful discussion. This work is supported by National Key R\&D Program of China No.2017YFA0304500, by National
Fundamental Research Program of China under Grant 2013CB921803, by National Natural
Science Foundation of China under grants No.91636220, No.91421111, No.11674360
and No.11375136, and by the research fund of City College, Wuhan University of
Science and Technology under grant No.2014CYBSKY001.
\end{acknowledgments}

\end{document}